\documentclass[letterpaper,twocolumn,prl,aps,superscriptaddress,floatfix]{revtex4}
\usepackage{mathrsfs}
\usepackage{mathptmx}

\usepackage[latin9]{inputenc}
\usepackage{amsmath}
\usepackage{graphicx}
\usepackage{esint}
\usepackage{diagbox}
\usepackage{multirow}
\usepackage{algpseudocode}
\usepackage{algorithm}

\makeatletter
\pdfpageheight\paperheight
\pdfpagewidth\paperwidth

\@ifundefined{textcolor}{}
{%
 \definecolor{BLACK}{gray}{0}
 \definecolor{WHITE}{gray}{1}
 \definecolor{RED}{rgb}{1,0,0}
 \definecolor{GREEN}{rgb}{0,1,0}
 \definecolor{BLUE}{rgb}{0,0,1}
 \definecolor{CYAN}{cmyk}{1,0,0,0}
 \definecolor{MAGENTA}{cmyk}{0,1,0,0}
 \definecolor{YELLOW}{cmyk}{0,0,1,0}
}
\usepackage{xcolor}
\usepackage{soul}

\newcommand{\bra}[1]{\ensuremath{\left\langle#1\right|}}
\newcommand{\ket}[1]{\ensuremath{\left|#1\right\rangle}}
\definecolor{blue}{rgb}{0,0,1}
\definecolor{red}{rgb}{1,0,0}
\definecolor{green}{rgb}{0,1,0}

\usepackage[unicode=true,pdfusetitle,
 bookmarks=true,bookmarksnumbered=false,bookmarksopen=false,
 breaklinks=false,pdfborder={0 0 0},pdfborderstyle={},backref=false,colorlinks=true]
 {hyperref}
\hypersetup{
 colorlinks,linkcolor=red,citecolor=blue}

\makeatother

\begin{document}
%\draft
\title{Experimental Quantum End-to-End Learning on a Superconducting Processor}

\author{X.~Pan}
\thanks{These two authors contributed equally to this work.}
\affiliation{Center for Quantum Information, Institute for Interdisciplinary Information Sciences, Tsinghua University, Beijing 100084, China}

\author {X.~Cao}
\thanks{These two authors contributed equally to this work.}
\affiliation{Center for Intelligent and Networked Systems, Department of Automation, Tsinghua University, Beijing 100084, China}

\author {W.~Wang}
\affiliation{Center for Quantum Information, Institute for Interdisciplinary Information Sciences, Tsinghua University, Beijing 100084, China}

\author {Z.~Hua}
\affiliation{Center for Quantum Information, Institute for Interdisciplinary Information Sciences, Tsinghua University, Beijing 100084, China}

\author {W.~Cai}
\affiliation{Center for Quantum Information, Institute for Interdisciplinary Information Sciences, Tsinghua University, Beijing 100084, China}

\author {X.~Li}
\affiliation{Center for Quantum Information, Institute for Interdisciplinary Information Sciences, Tsinghua University, Beijing 100084, China}

\author {H.~Wang}
\affiliation{Center for Quantum Information, Institute for Interdisciplinary Information Sciences, Tsinghua University, Beijing 100084, China}

\author {J.~Hu}
\affiliation{Center for Intelligent and Networked Systems, Department of Automation, Tsinghua University, Beijing 100084, China}

\author {Y.~P.~Song}
\affiliation{Center for Quantum Information, Institute for Interdisciplinary Information Sciences, Tsinghua University, Beijing 100084, China}

\author {Dong-Ling Deng}
\affiliation{Center for Quantum Information, Institute for Interdisciplinary Information Sciences, Tsinghua University, Beijing 100084, China}
\affiliation{Shanghai Qi Zhi Institute, No. 701 Yunjin Road, Xuhui District, Shanghai 200232, China}

\author{C.-L.~Zou}
\affiliation{Key Laboratory of Quantum Information, CAS, University of Science and Technology of China, Hefei, Anhui 230026, China}

\author{Re-Bing Wu}
\thanks{E-mail: rbwu@tsinghua.edu.cn}
\affiliation{Center for Intelligent and Networked Systems, Department of Automation, Tsinghua University, Beijing 100084, China}

\author{L.~Sun}
\thanks{E-mail: luyansun@tsinghua.edu.cn}
\affiliation{Center for Quantum Information, Institute for Interdisciplinary Information Sciences, Tsinghua University, Beijing 100084, China}

%\date{\today}

\begin{abstract}
Machine learning can be substantially powered by a quantum computer owing to its huge Hilbert space and inherent quantum parallelism. In the pursuit of quantum advantages for machine learning with noisy intermediate-scale quantum devices, it was proposed that the learning model can be designed in an end-to-end fashion, i.e., the quantum ansatz is parameterized by directly manipulable control pulses without circuit design and compilation. Such gate-free models are hardware friendly and can fully exploit limited quantum resources. Here, we report the first experimental realization of quantum end-to-end machine learning on a superconducting processor. The trained model can achieve $98\%$ recognition accuracy for two handwritten digits (via two qubits) and $89\%$ for four digits (via three qubits) in the MNIST (Mixed National Institute of Standards and Technology) database. The experimental results exhibit the great potential of quantum end-to-end learning for resolving complex real-world tasks when more qubits are available.
%\red{Machine learning can be substantially powered by quantum computation owing to the underlying huge Hilbert feature space and the inherent quantum parallelism.} To implement quantum machine learning algorithms on noisy intermediate-scale quantum devices, \red{the hardware-friendly ansatz that maximally exploits the limited quantum resources}. Here, we experimentally realize an \red{end-to-end machine learning task} on a superconducting platform, which parameterizes the ansatz with directly tunable quantum control parameters and a classical interface for encoding the input data to the control parameters. Using two or three qubits, the model trained by Adam algorithms can achieve $98\%$ recognition for two handwritten digits (two qubits) and $89\%$ for four digits (three qubits) in the MNIST (Mixed National Institute of Standards and Technology) database. The result shows that the end-to-end quantum machine models can be applied to real-world datasets with comparable performance achieved on much larger classical models.
\end{abstract}

\maketitle
\vskip 0.5cm
%\narrowtext

%\begin{eqnarray}
%H_{I} &=&-a_{1}^{\dagger }a_{1}\left( \chi_{11}\left\vert e\right\rangle
%_{1}\left\langle e\right\vert +\chi_{13}\left\vert e\right\rangle
%_{3}\left\langle e\right\vert \right) \notag\\
%&&-a_{2}^{\dagger }a_{2}\left( \chi_{22}\left\vert e\right\rangle
%_{2}\left\langle e\right\vert +\chi_{23}\left\vert e\right\rangle
%_{3}\left\langle e\right\vert \right) ,\mathscr{A}
%\label{equ:Ham}
%\end{eqnarray}

%\section{Introduction}
Quantum computing~\cite{nielsen2000quantum} is revolutionizing the field of machine learning (ML)~\cite{biamonte2017quantum,Dunjko2018,Sarma2019}. Powered by quantum Fourier transform and amplitude amplification, provable speed-up has been predicted for high-dimensional and big-data ML tasks using fault-tolerant quantum computers~\cite{lloyd2014quantum,rebentrost2014quantum,amin2018quantum,dunjko2016quantum,gao2018quantum}. Even with noisy intermediate-scale quantum (NISQ) devices, quantum advantage is still promising, because the model expressibility can be substantially enhanced by the exponentially large feature space carried by multi-qubit quantum states~\cite{liu2021rigorous,schuld2019quantum}.

To deploy quantum machine learning algorithms on NISQ processors, the key part is to construct a parameterized quantum ansatz that can be trained by a classical
optimizer. To date, most quantum ansatzes are realized by quantum neural networks (QNN)~\cite{havlivcek2019supervised,liu2021rigorous,benedetti2019parameterized,schuld2020circuit, farhi2018classification,wei2021quantum,houssein2021hybrid,farhi2014quantum,wurtz2021fixed,rudolph2020generation,zeng2019learning,jerbi2021variational} that consist of layers of parameterized quantum gates, and successful experiments have been demonstrated on classification~\cite{tacchino2019artificial,cai2015entanglement,johri2021nearest}, clustering~\cite{ouyang2020experimental,li2015experimental}, and generative~\cite{zoufal2019quantum,hu2019quantum,Zhu2019} learning tasks. The gate-based QNN ansatz naturally incorporates the theory of quantum circuits, but the learning performance is highly dependent on the architecture design and the mapping of circuits to experimentally operable native gates. A structurally non-optimized QNN cannot fully exploit the limited quantum coherence resource, and this is partially why high learning accuracy is hard to attain on NISQ devices without downsizing the training dataset.

There are certainly much room for performance improvement by using more hardware-efficient quantum ansatzes, e.g., via deep optimization of the circuit architecture~\cite{ostaszewski2021reinforcement} and qubit mapping strategies~\cite{Huang2021PowerOD}. Recently, a hardware-friendly end-to-end learning scheme (in the sense that the model is trained as a whole instead of being divided into separate modules) is proposed~\cite{Wu2020} by replacing the gate-based QNN with natural quantum dynamics driven by coherent control pulses. This model requires very little architecture design, system calibration, and no qubit mapping. One can also jointly train a data encoder that automatically transforms classical data to quantum states via control pulses, and this essentially simplifies the encoding process because the preparation of quantum states according to a hand-designed encoding scheme is no more required. More importantly, the natural control-to-state mapping involved in the encoding process introduces nonlinearity that is crucial for better model expressibility.

In this paper, we report the first experimental demonstration of quantum end-to-end machine learning using a superconducting processor through the recognition of handwritten digits selected from the MNIST (Mixed National Institute of Standards and Technology) dataset. Without downsizing the original 784-pixel images, the end-to-end learning model can be trained to achieve $98\%$ accuracy with two qubits for the 2-digit classification and $89\%$ accuracy with three qubits for the 4-digit task, which are among the best experimental results reported on small-size quantum processors~\cite{Wang2021ExperimentalRO}. The demonstrated quantum end-to-end model can be easily scaled up for solving complex real-world learning tasks owing to its inherent hardware friendliness and efficiency.

%In this paper, we demonstrate the applied end-to-end quantum machine learning algorithm in a superconducting circuit to classify handwritten digits in the MNIST database. \red{Using two or three qubits, the model trained by Adam algorithms can achieve $98\%$ recognition for two handwritten digits (two qubits) and $89\%$ for four digits (three qubits) in the MNIST (Mixed National Institute of Standards and Technology) database. The result shows that quantum machine models can be applied to real-world datasets with comparable performance achieved on much larger classical models.}

\begin{figure*}
\includegraphics[scale=1]{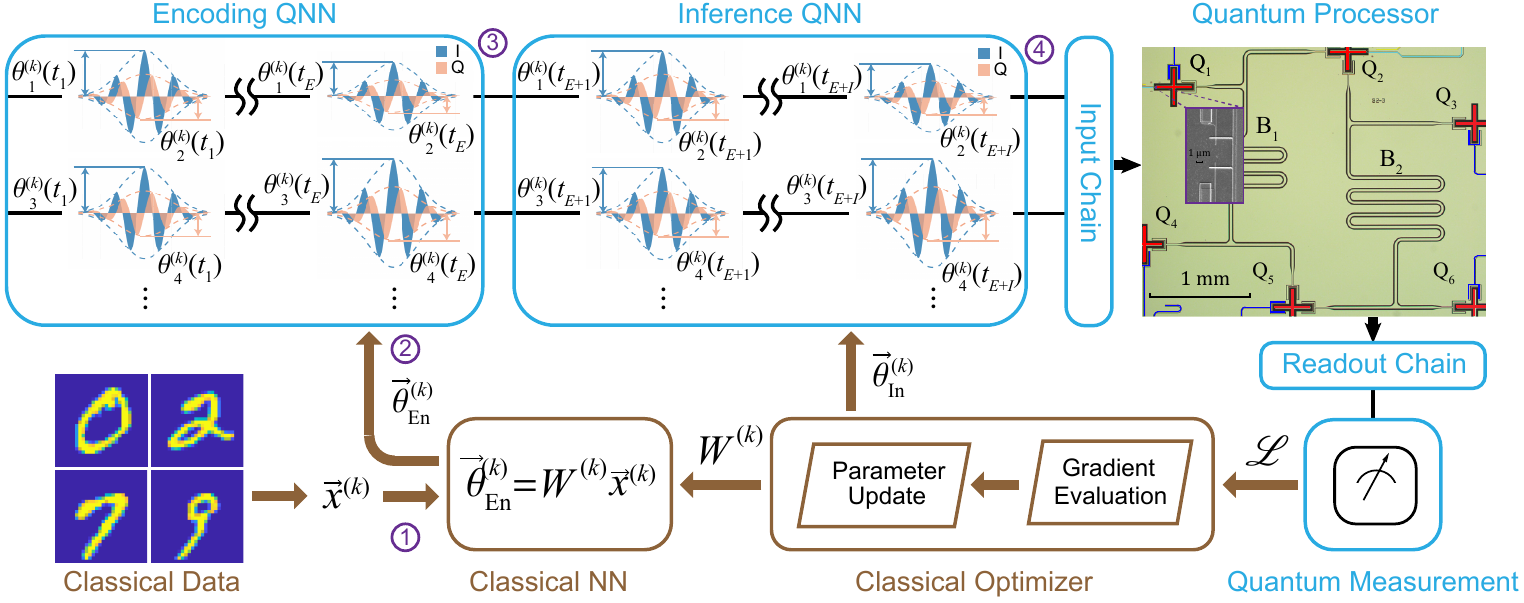}
\caption{(Color online) The training protocol of the quantum end-to-end learning framework. In the $k$-th iteration, a randomly selected image of a handwritten digit in the MNIST dataset is converted to a vector $\vec{x}^{(k)}$ and then transformed by a matrix $W^{(k)}$ to the control variables $\vec{\theta}^{(k)}_\mathrm{En}$ that steer the quantum state to $\ket{\psi^{(k)}(t_E)}$ of the qubits in the QNN. This process encodes $\vec{x}^{(k)}$ to $\ket{\psi^{(k)}}$. Subsequent inference control pulses $\vec{\theta}^{(k)}_\mathrm{In}$ are applied to drive $\ket{\psi^{(k)}(t_E)}$ to $\ket{\psi^{(k)}(t_{E+I})}$ that is to be measured. The parameters in $W^{(k)}$ and $\vec{\theta}^{(k)}_\mathrm{In}$ are updated for the next iteration according to the loss function $\mathcal{L}$ and its gradient obtained from the measurement. The circled numbers represent the specific points in the data flow and the corresponding learning performances are shown in Fig.~\ref{Fig:LDA}. The top right is the false-colored optical image of the six-qubit device used in our experiment.}
%The encoding layer parameters $\vec{\theta}^{(k)}_\mathrm{En}$ and the inference layer parameters $\vec{\theta}^{(k)}_\mathrm{In}$ are directly sent to the arbitrary waveform generator (AWG) for the corresponding qubit controls. The inference layer are independent of the input data and are introduced for increasing the depth and complexity of the QNN. The measured distribution in the computational bases spanned by the output qubits are sent to a classical optimizer. The parameters in $W^{(k)}$ and $\vec{\theta}^{(k)}_\mathrm{In}$ are then updated according to the loss function and the gradient obtained from the output of the QNN.
%\blue{We note that the on-chip wire bonding process that is used in the this work is not included for simplification and the complete experimental setup can be found in the Appendix. The two large winding structures without coloring are the coupling buses. The shunting capacitors of the transmon qubits are colored red, while the SEM image of the qubit junctions is shown in the inset. The readout resonators, flux bias lines, and qubit RF drive lines are colored dark blue, light blue, and yellow respectively.}
\label{Fig:fig1}
\end{figure*}

The basic idea of end-to-end quantum learning is to parameterize the quantum ansatz by physical control pulses that are usually applied to implement abstract quantum gates in variational quantum classifiers. In this way, a feedforward QNN can be constructed by the control-driven evolution of the quantum state $\ket{\psi(t)}$, as follows~\cite{rivas2012open}:
\begin{equation}
\frac{d\ket{\psi(t)}}{dt}=-\frac{i}{\hbar}\left[H_0+\sum_{m=1}^{r} \theta_m(t) H_m\right]\ket{\psi(t)},
\label{equ:Hamiltonian}
\end{equation}
where $H_0$ is the static Hamiltonian which involves the coupling between different qubits, and $r$ is the number of pulsed control functions/channels in the quantum processor. For example, if there are $M$ qubits for the QNN and each qubit is dictated by $c$ control functions (e.g., flux bias or microwave driving), we have $r=c\times M$. Here, $H_m$ is the control Hamiltonian associated with the $m$-th control pulse that contains $n$ sub-pulses over $n$ sampling periods. The $j$-th sub-pulse is parameterized by $\theta_m(t_j)$, and hence we denote the $m$-th control pulse by $\vec{\theta}_m=[{\theta}_m(t_1),{\theta}_m(t_2),...,{\theta}_m(t_n)]$. The evolution of the quantum system under all $n$-th control sub-pulses constructs the $n$-th layer of the QNN.

We illustrate the quantum end-to-end learning with a classification task based on the MNIST dataset. As shown in Fig.~\ref{Fig:fig1}, an image of a handwritten digit is randomly selected from the training dataset $\mathcal{D}$. In the $k$-th iteration, the sampled image is converted to a $d=784$ dimensional vector $\vec{x}^{(k)}$, and $y^{(k)}$ is the corresponding label. The input data $\vec{x}^{(k)}$ is transformed by a matrix $W^{(k)}$ to the control variables $\vec{\theta}^{(k)}_\mathrm{En}=W^{(k)}\vec{x}^{(k)}$. This constructs a classical encoding block with $r$ channels and $E$ sub-pulses per channel: $\vec{\theta}^{(k)}_\mathrm{En} =[{\theta}^{(k)}_1(t_1),...,{\theta}^{(k)}_1(t_\mathrm{E}),...,
{\theta}^{(k)}_r(t_1)...,{\theta}^{(k)}_r(t_E)]$. The generated control pulses $\vec{\theta}^{(k)}_\mathrm{En}$ then automatically encode $\vec{x}^{(k)}$ to the quantum state $\ket{\psi^{(k)}(t_E)}$ via the natural quantum state evolution of Eq.~(\ref{equ:Hamiltonian}).

Subsequent inference control pulses $\vec{\theta}^{(k)}_\mathrm{In}$, which have the same form as $\vec{\theta}^{(k)}_\mathrm{En}$ but consist of $I$ sub-pulses in each channel, are then applied to induce the quantum evolution from the encoded quantum state $\ket{\psi^{(k)}(t_E)}$. The inference controls are introduced for improving the classification performance. Finally, the end-time quantum state $\ket{\psi^{(k)}(t_{E+I})}$ is measured under the appropriate experiment-available positive operator $O^{(k)}$ according to the classical label $y^{(k)}$, which gives the conditional probability (or confidence) of obtaining $y^{(k)}$ for a given input $\vec{x}^{(k)}$
\begin{equation}
P(y^{(k)}|\vec{x}^{(k)},W^{(k)},\vec{\theta}^{(k)}_\mathrm{{In}})=\bra{\psi^{(k)}(t_{E+I})}O^{(k)}\ket{\psi^{(k)}(t_{E+I})}. \label{equ:confidence}
\end{equation}
The corresponding loss function is defined as
\begin{equation}
\mathcal{L}[W^{(k)},\vec{\theta}^{(k)}_\mathrm{In},\{\vec{x}_{\ell}^{(k)}\}]=1-\frac{1}{b}\sum_{\ell=1}^bP(y^{(k)}|\vec{x}_\ell^{(k)},W^{(k)},\vec{\theta}^{(k)}_\mathrm{{In}}).
\label{equ:loss}
\end{equation}
In the experiment, we select a batch of $b$ samples in each iteration to reduce the fluctuation of $\mathcal{L}$ for faster convergence of the learning process. The gradient of the loss function $\mathcal{L}$ with respect to the encoding control $\vec{\theta}^{(k)}_\mathrm{En}$ and the inference control $\vec{\theta}^{(k)}_\mathrm{In}$ can be evaluated with the finite difference method by making a small change of the each control parameter $\theta^{(k)}_i(t_j)$~\cite{li2005general}. The gradient of $\mathcal{L}$ with respect to $W^{(k)}$ can be derived from the gradient of $\mathcal{L}$ with respect to $\vec{\theta}^{(k)}_\mathrm{En}$~\cite{Wu2020}. Therefore, we can apply the widely used stochastic gradient-descent algorithm in machine learning to update $W^{(k)}$ and $\vec{\theta}^{(k)}_\mathrm{In}$ by minimizing $\mathcal{L}$ on the training dataset $\mathcal{D}$ (see Supplementary Materials for details of the algorithms)~\cite{friedman2002stochastic}. Once the model is well trained, one can use fresh samples from a testing dataset to examine the recognition performance of the handwritten digits.

%Note that in order to make $\mathcal{L}$ have a smaller fluctuation for a better convergence of the learning process, one can use a batch of input data with a batch size of $b$, instead of one single $\vec{x}^{(k)}$, with the same $W^{(k)}$ and $\vec{\theta}^{(k)}_\mathrm{In}$ to get the averaged $\mathcal{L}$  \blue{and its gradient} for each round, which is also referred as an epoch corresponding to each update of $W^{(k)}$ and $\vec{\theta}^{(k)}_\mathrm{In}$.

\begin{figure}
    \includegraphics[scale=1]{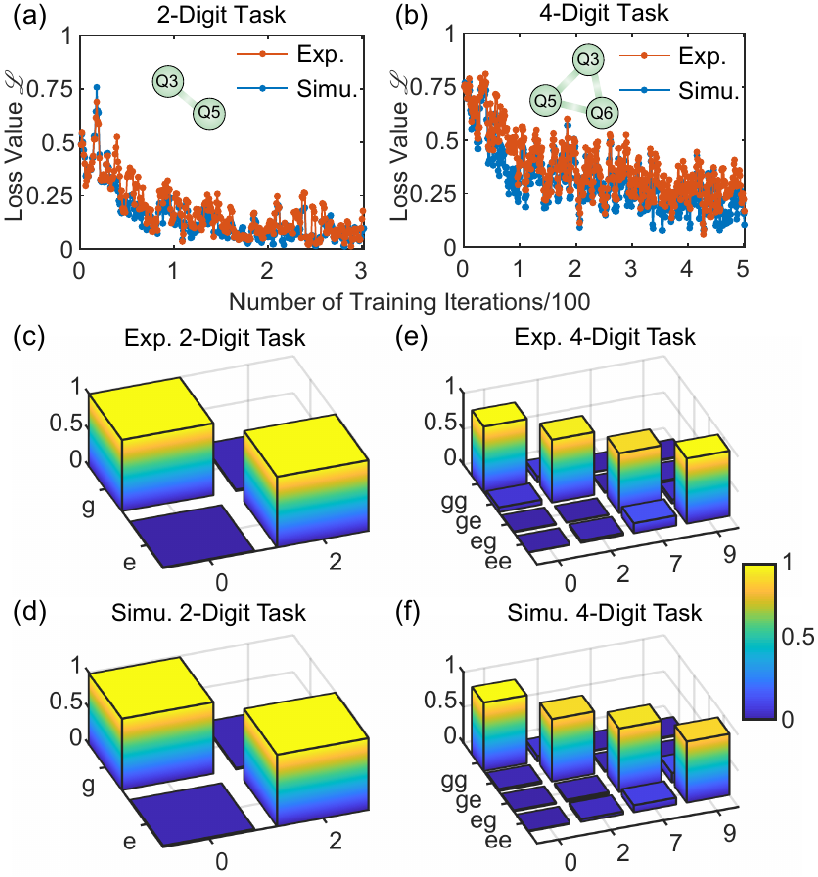}
    \caption{(Color online) (a) and (b) The typical training process of the end-to-end model. For better clarity, all data points are averaged over the neighboring four points. (c-f) The classification performance of the trained model. The horizontal labels show the digits to be classified, while the vertical labels show the majority vote of the computational basis measurement results. (c) and (d) Experimental and simulation results for the 2-digit classification task, respectively. The averaged accuracies for the classification are $0.987$ and $0.982$ in the experiment and the simulation, respectively. (e) and (f) The 4-digit classification of the QNN. The averaged accuracies are 0.895 and 0.889 in the experiment and the simulation, respectively.}
\label{Fig:fig2}
   %\red{The cross section means the probability that the test data labeled by `0' or `2' is classified as `g' (`0') or `e' (`2').} If the classification is perfect, the diagonal elements of the probability matrix will be unity.
\end{figure}

The end-to-end model is demonstrated in a superconducting processor, as shown in Fig.~\ref{Fig:fig1}. All qubits take the form of the flux-tunable Xmon geometry and are driven with inductively coupled flux bias lines and capacitively coupled RF control lines~\cite{Barends2013,Li2018,Cai2019}. Among the six qubits, $Q_1,Q_2,Q_4,Q_5$ are dispersively coupled to a half-wavelength coplanar cavity $B_1$, and $Q_2,Q_3,Q_5,Q_6$ are dispersively coupled to another cavity $B_2$. Each qubit is dispersively coupled to a quarter-wavelength readout resonator for a high-fidelity single-shot readout and all the resonators are coupled to a common transmission line for multiplexed readouts. The qubits that are not relevant to the QNN is biased far away and can be ignored from the system Hamiltonian, therefore, the static Hamiltonian of the QNN can be written in the interaction picture as
\begin{eqnarray}
H_0/\hbar=\sum_{q\neq{p}}J_{qp}(a_q^\dagger{}a_p+a_p^\dagger{}a_q)
-\sum_{q=1}^M{}\frac{E_{C,q}}{2}a_q^\dagger{}a_q^\dagger{}a_qa_q,
\label{equ:H_0}
\end{eqnarray}
where $J_{qp}$ is the coupling strength between the $p$-th and $q$-th qubits mediated by the bus cavity, $E_{C,q}$ denotes the qubit anharmonicity, and $a_q$ is the annihilation operator of the $q$-th qubit.

%Qubits (the $p$-th and $q$-th ones) that couple to the common cavity share a cavity-mediated coupling strength $J_{qp}=g_{qB}g_{pB}(\Delta_{qB}+\Delta_{pB})/2\Delta_{qB}\Delta_{pB}$~\cite{Song2017}. Here $g_{qB}$ and $\Delta_{qB}$ denote the coupling strength and the detuning between $Q_q$ and coupling cavity respectively.

Throughout this work, we set the encoding block with $E=2$ layers followed by an inference block with $I=2$ layers. As shown in Fig.~\ref{Fig:fig1}, for the $q$-th qubit in the $n$-th ($n=1,2,3,4$) layer of the QNN, there are $c=2$ control parameters $\theta_{2q-1}(t_{n})$ and $\theta_{2q}(t_{n})$, which are associated with the control Hamiltonians $H_{2q-1}=(a_q+a_q^\dag)/2$ (rotation along the $x$-axis of the Bloch sphere) and $\quad H_{2q}=(ia_q-ia_q^\dag)/2$ (rotation along the $y$-axis of the Bloch sphere), respectively. The control parameters are the variable amplitudes of the Gaussian envelopes of two resonant microwave sub-pulses, each of which has a fixed width of $4\sigma=40$~ns. All the quantum controls in the same time interval are exerted simultaneously. For an $N$-digit classification task, we take $M=\lceil \log_2{N} \rceil+1$ qubits for the QNN: the classification results are mapped to the computation bases of the first $\lceil \log_2{N} \rceil$ qubits (label qubits) by the majority vote of the collective measurement performed on label qubits, while one additional qubit is introduced for a better expressibility of the model. Therefore, the QNN in our experiment involves totally $cM(E+I)=8M$ control parameters.

%\red{To summarize, we can use the controlled quantum dynamics to construct a QNN model, in which the total control Hamiltonian in the $n$-th layer can be written as
%    \begin{eqnarray}
%    H_c(t_n)=\sum\limits_{q=1}^{M}\left(\frac{\theta_{2q-1}(t_n)(a_q+a_q^\dagger)}{2}
%    +\frac{\theta_{2q}(t_n)(ia_q-ia_q^\dagger)}{2}\right).
%    \label{equ:H_ell}
    %U_\ell=\prod_{j=1}^{N}\exp(-\frac{i\theta_{2\ell-1,j}\sigma_{x,j}}{2}
    %-\frac{i\theta_{2\ell,j}\sigma_{y,j}}{2}),
    %\label{equ:U_ell}
%    \end{eqnarray}
%    The QNN model involves totally $2M(E+I)$ parameters.}

%\section{Results}
%\subsection{The Training Protocol of the end-to-end model}

We perform the 2-digit (`0' and `2') classification task ($N=2$) with $Q_3$ and $Q_5$ ($M=2$). The working frequencies are $6.08$~GHz and $6.45$~GHz, respectively, which are also the flux sweet-spots of the two qubits. The effective coupling strength $J_{35}/2\pi=4.11$~MHz. We take $Q_5$ as the label qubit and assign the classification result to be `0' or `2' if the respective probability of measuring $\ket{g}$ or $\ket{e}$ state is larger.

The end-to-end model is initialized with $W=W_0$ and $\vec{\theta}_\mathrm{In}=\vec{\theta}_0$, where all elements of $W_0$ are $10^{-5}$ and each element of $\vec{\theta}_0$ is tuned to induce a $\pi/4$ rotation of the respective qubit. The parameter update is realized as follows. Firstly, we obtain the loss function $\mathcal{L}$ according to Eq.~(\ref{equ:loss}) by measuring $Q_5$. We perturb each control parameter in the control set $\{\vec{\theta}_\mathrm{En},\vec{\theta}_\mathrm{In}\}$ and obtain the corresponding gradient of $\mathcal{L}$. The $\mathcal{L}$ and its gradient averaged over a batch of two training samples ($b=2$) are sent to a classical Adam optimizer~\cite{Kingma2015} for updating $W$ and $\vec{\theta}_\mathrm{In}$. All control parameters are linearly scaled to the digital-to-analog converter level of a Tektronix arbitrary waveform generator 70002A, working with a sampling rate of $25$ GHz, to generate the resonant RF pulses directly. The control pulses composed of in-phase and quadrature components are sent to each qubit with the corresponding RF control line. To obtain the classification result, we repeat the procedure and measure the label qubit for $5000$ times.

In the 4-digit ('0', '2', '7', and '9') classification task ($N=4$), we take $Q_3$, $Q_5$, and $Q_6$ ($M=3$) to construct the QNN, whose working frequencies are 6.08~GHz, 6.45~GHz, and 6.19~GHz, respectively. $Q_3$ and $Q_5$ are measured for the classification output. The target digits correspond to the four computational bases spanned by the two label qubits. The training procedure and algorithms are the same as those for the $N=2$ task.

%The numerical simulation of the system dynamics is performed also with the calibrated open system Hamiltonian and the same training dataset, as well as the parameter update algorithm that is used in the experiment.

The typical training process is shown in Figs.~\ref{Fig:fig2}a-b. For better clarity, the curves are smoothed out by averaging each data point from its neighboring four ones. For the 2-digit (4-digit) classification task, the experimental loss function $\mathcal{L}$ converges to 0.14 (0.22) in 300 (500) iterations. The training loss can potentially be reduced by increasing the depth $E$ of the encoding block~\cite{Wu2020}. For comparison, numerical simulations are also performed with the calibrated system Hamiltonian, the same batches of training samples, and the same parameter update algorithms. As shown in Figs.~\ref{Fig:fig2}a-b, the simulations match the experiments well. The small deviation of the experimental data may attribute to the simplified modeling of high-order couplings between the qubits and the control pulses~\cite{Motzoi2009}, as well as the system parameter drifting.

%\blue{However, in machine learning standards, the performance of a trained network is more significant than the calibration preciseness, which becomes unaffordable when the system size and geometry complexity grow.}

%\subsection{Benchmark the trained end-to-end model}

To examine the performance of the end-to-end learning, we experimentally test the generalizability of the trained end-to-end model with fresh testing samples (1000 for each digit), and count the frequencies of assigning these samples to different digits (see Figs.~\ref{Fig:fig2}c-f). The measured overall accuracies (i.e., the proportion of samples that are correctly classified) are $98.7\%$ for the 2-digit task and $89.5\%$ for the 4-digit task, respectively, which are consistent with the simulation results ($98.2\%$ and $88.9\%$, respectively) based on the experimentally identified Hamiltonian.

\begin{figure}
    \includegraphics[scale=1]{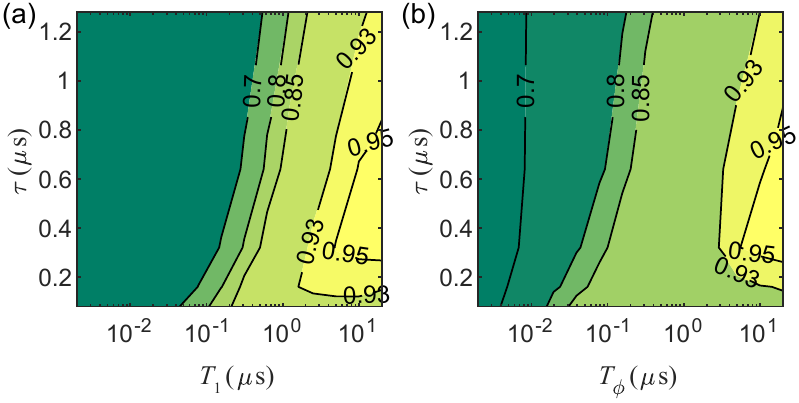}
    \caption{(a) The simulated average confidence as a function of the qubit relaxation time $T_1$ and the pulse length $\tau$, while the qubit pure dephasing time $T_\phi$ is infinite; (b) The simulated average confidence as a function of $T_\phi$ and $\tau$, while $T_1$ is infinite.}
    \label{fig3}
\end{figure}

The performance of the model also relies on the amount of entanglement gained in the quantum state. When the number of QNN layers is fixed, the quantum state gets more entangled under longer pulse length $\tau$ (includes all the $E+I=4$ sub-pulses in both the encoding and the inference blocks), but coherence may be lost in the prolonged control time duration due to the inevitable decoherence. We use the experimentally calibrated parameters to simulate the 2-digit classification process under different $\tau$ and different coherence times $T_1$ and $T_\phi$ of the qubits. As shown in Fig.~\ref{fig3}, the average confidence $1-\mathcal{L}$ varies little with $\tau$ when $T_1$ or $T_\phi$ is sufficiently small because the coherent control is overwhelmed by the strong decoherence. For larger $T_1$ or $T_\phi$ (e.g., $T_1=20~\mu$s), the average confidence initially increases with $\tau$, but then decreases after reaching the peak. This trend clearly indicates the trade-off between the gained entanglement and the lost coherence, and thus $\tau$ as well as the number of layers should be optimally chosen for the best balance.

The end-to-end learning scheme provides a seamless combination of quantum and classical computers through the joint training of the control-based QNN and the classical data encoder $W$. To understand their respective roles in the classification, we check how the data distribution varies along the flow $\vec{x}\rightarrow\vec{\theta}_\mathrm{En}\rightarrow\ket{\psi(t_{E})}\rightarrow\ket{\psi(t_{E+I})}\rightarrow y$ (see $\textcircled{1}\sim\textcircled{4}$ in Fig.~\ref{Fig:fig1}) in the 2-digit classification process. To facilitate the analysis, we use the Linear Discriminant Analysis (LDA)~\cite{venables2013modern} that projects high-dimensional data vectors into two clusters of points distributed on an optimally chosen line (see details in the Supplementary Materials). The LDA makes it easier to visualize and compare data distributions whose dimensionalities are different.

%we check the data distribution along the flow $\vec{x}\rightarrow\vec{\theta}_\mathrm{En}\rightarrow\ket{\psi(t_{E})}\rightarrow\ket{\psi(t_{E+I})}\rightarrow y$ (see $\textcircled{1}\sim\textcircled{4}$ in Fig.~\ref{Fig:fig1}) in the 2-digit classification via the Linear Discriminant Analysis (LDA)~\cite{venables2013modern}. The LDA projects the high-dimensional data vectors into two clusters of points distributed on an optimally chosen line so that the two clusters are separated as far as possible (see details in the Supplementary Materials).

%We apply the LDA analysis along the data flow from the original two-digit dataset (\textcircled{1}) to the transformed control variables (\textcircled{2}), and then to the encoded quantum state (\textcircled{3}) and the final output state (\textcircled{4}).

\begin{figure}
    \includegraphics[scale=1]{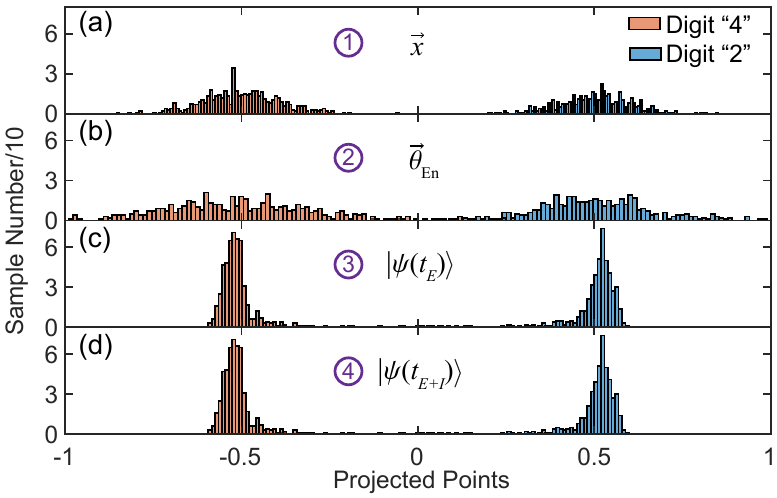}
    \caption{The projected points by using LDA, where the distance between the projected cluster centers is normalized. The positions of $\textcircled{1}\sim\textcircled{4}$ along the data flow are shown in Fig.~\ref{Fig:fig1}. (a) The projected points of the original 784-dimensional handwritten digits. (b) The projected points of the control pulses for encoding the classical data to the quantum states. (c) The projected points of the quantum states after the encoding block. (d) The projected points of the final quantum states.}
    \label{Fig:LDA}
\end{figure}

The projected clusters are plotted in Fig.~\ref{Fig:LDA}. In each sub-figure, the distance between the centers of the two clusters is normalized, and hence we can quantify the classifiability by their standard deviations (i.e., the narrowness of distribution). As can be seen in Figs.~\ref{Fig:LDA}a-b, the classical data encoder $W$ effectively reduces the original 784-dimensional vector $\vec{x}$ to a 8-dimensional vector of control variables $\vec{\theta}_\mathrm{En}$, but the standard deviation is increased from $0.1658$ for the original dataset to $0.2903$ for the transformed control pulses. Then, the control-to-state mapping, which is both nonlinear and quantum, sharply reduces the standard deviation to $0.0919$ for the encoded quantum state (Fig.~\ref{Fig:LDA}c), while the following quantum inference block does not make further improvement (Fig.~\ref{Fig:LDA}d). These results indicate that the classical data encoder is responsible for the compression of the high-dimensional input data, while the classification is mainly accomplished by the QNN.

%\section{Discussion and Summary}
It should be noted that no quantum advantage is claimed here with a small-size NISQ processor. However, we notice that the end-to-end learning is very similar to quantum reservoir computing (QRC)~\cite{mujal2021opportunities,ghosh2019quantum} in that both schemes exploit complex natural quantum dynamics for hard computing tasks, and QRC has been proven to have universal approximation property~\cite{chen2020temporal} and higher information processing capacity~\cite{bravo2021quantum}. It is conjectured that similar conclusions can be made for quantum end-to-end learning, and these will be explored in our future studies.

Apparently, the power of QNN can be exponentially increased when more controllable qubits are available. We expect to further improve the training efficiency of the end-to-end learnings on larger NISQ processors, and develop more complicated ML application (e.g., unsupervised and generative learning) based on more complex datasets.

This work was supported by National Key Research and Development Program of China (Grants No.~2017YFA0304303 and 2017YFA0304304), the National Natural Science Foundation of China (Grants No.~92165209, No.~61833010, No.~62173201, No.~11925404, No.~11874235, No.~11874342, No.~11922411, No.~12061131011, No.~12075128), Key-Area Research and Development Program of Guangdong Provice (Grant No.~2020B0303030001), Anhui Initiative in Quantum Information Technologies (AHY130200), China Postdoctoral Science Foundation (BX2021167), and Grant No.~2019GQG1024 from the Institute for Guo Qiang, Tsinghua University. D.-L. D. also acknowledges additional support from the Shanghai Qi Zhi Institute.

%\bibliographystyle{Zou}
%\bibliography{citation}

%Control: production of eprint (-1) disabled
%

\end{document}

\draft
\title{Supplementary Materials for ``Experimental Quantum End-to-End Learning on a Superconducting Processor"}

\author{X.~Pan}
\thanks{These two authors contributed equally to this work.}
\affiliation{Center for Quantum Information, Institute for Interdisciplinary Information Sciences, Tsinghua University, Beijing 100084, China}

\author {X.~Cao}
\thanks{These two authors contributed equally to this work.}
\affiliation{Center for Intelligent and Networked Systems, Department of Automation, Tsinghua University, Beijing 100084, China}

\author {W.~Wang}
\affiliation{Center for Quantum Information, Institute for Interdisciplinary Information Sciences, Tsinghua University, Beijing 100084, China}

\author {Z.~Hua}
\affiliation{Center for Quantum Information, Institute for Interdisciplinary Information Sciences, Tsinghua University, Beijing 100084, China}

\author {W.~Cai}
\affiliation{Center for Quantum Information, Institute for Interdisciplinary Information Sciences, Tsinghua University, Beijing 100084, China}

\author {X.~Li}
\affiliation{Center for Quantum Information, Institute for Interdisciplinary Information Sciences, Tsinghua University, Beijing 100084, China}

\author {H.~Wang}
\affiliation{Center for Quantum Information, Institute for Interdisciplinary Information Sciences, Tsinghua University, Beijing 100084, China}

\author {J.~Hu}
\affiliation{Center for Intelligent and Networked Systems, Department of Automation, Tsinghua University, Beijing 100084, China}

\author {Y.~P.~Song}
\affiliation{Center for Quantum Information, Institute for Interdisciplinary Information Sciences, Tsinghua University, Beijing 100084, China}

\author {Dong-Ling Deng}
\affiliation{Center for Quantum Information, Institute for Interdisciplinary Information Sciences, Tsinghua University, Beijing 100084, China}
\affiliation{Shanghai Qi Zhi Institute, No. 701 Yunjin Road, Xuhui District, Shanghai 200232, China}

\author{C.-L.~Zou}
\affiliation{Key Laboratory of Quantum Information, CAS, University of Science and Technology of China, Hefei, Anhui 230026, China}

\author{Re-Bing Wu}
\thanks{E-mail: rbwu@tsinghua.edu.cn}
\affiliation{Center for Intelligent and Networked Systems, Department of Automation, Tsinghua University, Beijing 100084, China}

\author{L.~Sun}
\thanks{E-mail: luyansun@tsinghua.edu.cn}
\affiliation{Center for Quantum Information, Institute for Interdisciplinary Information Sciences, Tsinghua University, Beijing 100084, China}

%\date{\today}
\maketitle

\section{Device Fabrication and Device Parameters}

The device is fabricated on a 2-inch $c$-plane sapphire substrate. A 100-nm aluminum film is first directly deposited in Plassys MEB 550S without any precleaning treatment. The resonators and control lines are then defined with photolithography followed by inductively-coupled-plasma etching. The wafer is diced into $8$mm$\times8$mm chips for the subsequent process. Double-angle evaporations are applied to define the qubit junctions. All control signals are transmitted through an SMA connector, a welded printed circuit board, and wire-bonding lines to the chip. The marginal ground plane of the chip is densely bonded to the aluminum sample box. On-chip wire-bondings are also applied to connect different central ground areas on the chip to suppress the spurious slot-line modes.

For clarity, we show in Fig.~1 of the main text the device that is similar to the one used in our experiment, but without the on-chip wire bondings. All qubits ($Q_1-Q_6$) are flux-tunable Xmon qubits~\cite{Barends2013,Li2018,Cai2019} with individual driving lines and flux bias lines. Each qubit is dispersively coupled to a quarter-wavelength readout resonator for a high-fidelity single-shot readout and all the readout resonators are coupled to a common transmission line for multiplexed readouts. Among the six qubits, $Q_1,Q_2,Q_4,Q_5$ are dispersively coupled to a half-wavelength coplanar cavity $B_1$, and $Q_2,Q_3,Q_5,Q_6$ are dispersively coupled to another cavity $B_2$. Qubits (the $p$-th and $q$-th ones) that couple to the common cavity share a cavity-mediated coupling strength $J_{qp}=g_{qB}g_{pB}(\Delta_{qB}+\Delta_{pB})/2\Delta_{qB}\Delta_{pB}$~\cite{Song2017}. Here $g_{qB}$ and $\Delta_{qB}$ denote the coupling strength and the detuning between $Q_q$ and the bus cavity, respectively. The parameters of the qubits and the bus cavity that are used in the experiment are shown in Table~\ref{qubit-cal}.

\begin{table}
\begin{tabular}{c c c c c}
  \hline
  \hline
  parameter & $Q_3$ & $Q_5$ & $Q_6$ & $B_1$\\
  \hline
  frequency (GHz) (sweet spot) & 6.081 & 6.453 & 6.191 & 5.90 \\
  \hline
  anharmonicity (MHz) & 217 & 206 & 197 & \\
  \hline
  \multirow{3}*{coupling strength $J_{pq}/2\pi$ (MHz)} & {4.11} & 4.11 & $\diagdown$ \\ \cline{2-4}
   & $\diagdown$ & {3.02} & 3.02 \\ \cline{2-4}
   & 4.46 & $\diagdown$ & 4.46 & \\
  \hline\\
  coupling strength $g_{pB}/2\pi$ (MHz) & 31 & 34 & 32 &  \\
  \hline\\
  energy relaxation time $T_1$ ($\mu$s) & 13 & 10 & 13 & \\
  \hline
  pure dephasing time $T_\phi$ ($\mu$s) & 49 & 37 & 58 & \\
  \hline
  \hline
\end{tabular}
\caption{Parameters of the qubits and the bus cavity that are used in the experiment.}
\label{qubit-cal}
\end{table}

\section{Experimental Setup}

\begin{figure*}
    \includegraphics{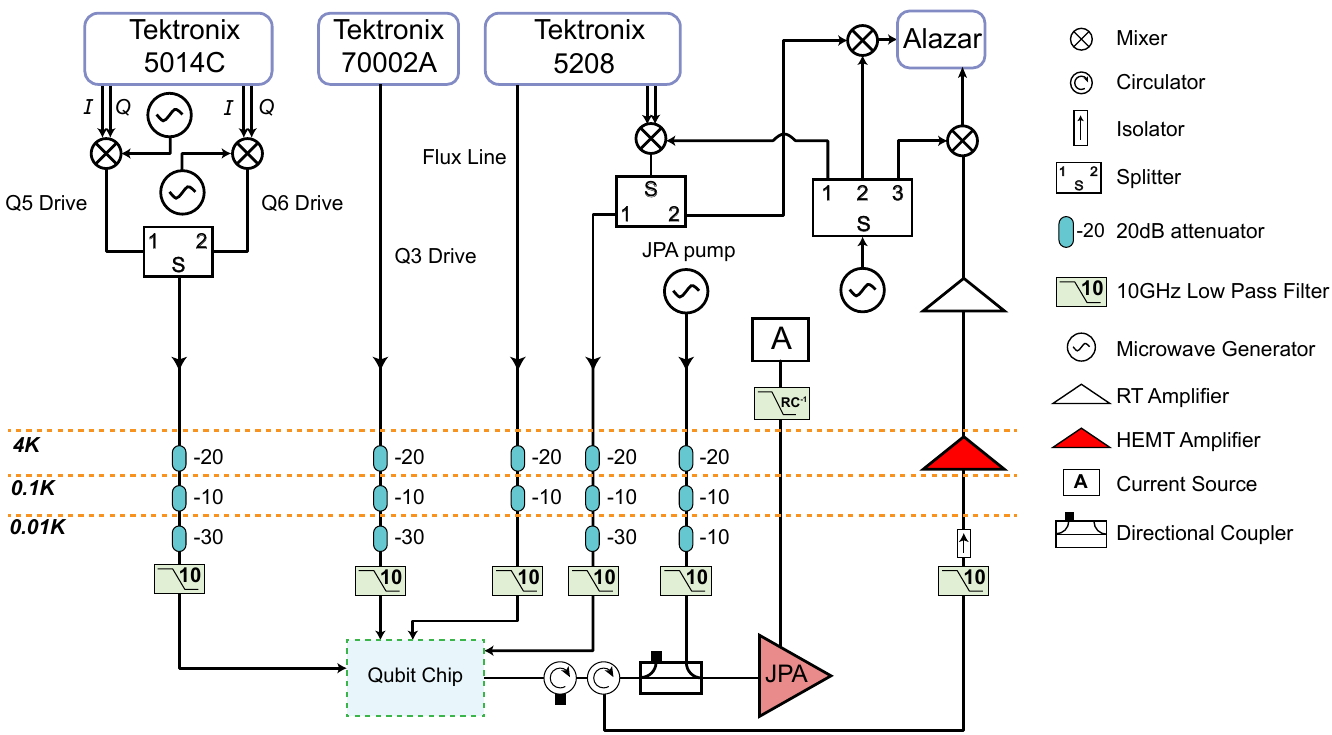}
    \caption{Experimental wiring.}
    \label{Fig:Setup}
\end{figure*}

The assembled aluminum sample box is fixed to a copper plate that is anchored to the 10~mK base plate of a dilution refrigerator. Additional copper braiding lines between the sample box and the copper plate are used for a better thermal connection. The sample box is also enclosed within a cylindrical magnetic shield. The measurement wiring outside the magnetic shield is shown in Fig.~\ref{Fig:Setup}.

The control signals for high-frequency drives are properly attenuated and low-pass filtered below 10~GHz. The flux bias lines for tuning qubit frequencies are low-pass filtered below 1~GHz. The input readout signals are generated from the same local microwave source and modulated with different frequency tones provided by a Tektronix AWG5208. The readout signals are multiplexed through a transmission line that couples to all the six readout resonators. The output readout signal is amplified with a Josephson parametric amplifier (JPA)~\cite{Hatridge,Roy2015} working at the base temperature. A high-electron-mobility-transistor amplifier working at 4~K temperature and an additional room temperature amplifier are also used to improve the signal-to-noise ratio before the detection with the digitizer at room temperature.

\section{Readout of Device}

\begin{table}
\begin{tabular}{c c c c}
  \hline
  \hline
  & $R_3$ & $R_5$ & $R_6$ \\
  \hline
  frequency (GHz) & 7.12 & 7.22 & 7.21 \\
  \hline
  dispersive shift (MHz) & 2.0 & 2.3 & 2.0 \\
  \hline
  Readout resonator lifetime (ns) & 150 & 110 & 160\\
  \hline
  Readout time (ns) & 800 & 800 &  \\
  \hline
  \hline
\end{tabular}
\caption{Readout parameters.}
\label{readout-cal}
\end{table}

The readout signal of each qubit is obtained by digitizing the corresponding down-converted output signal. The ground state $\ket{g}$ or the excited state $\ket{e}$ of the qubit is discriminated by taking a specific threshold value for each run of the measurement. The readout parameters of the three qubits that are used in this work are shown Table~\ref{readout-cal}.

We apply a Bayesian correction to the readout outcome to suppress the qubit decay errors and the qubit state discrimination errors due to insufficient separation of the $\ket{g}$ and $\ket{e}$ states. Therefore, the final probability distribution of the computational bases as a column vector $\vec{P}^\prime$ can be obtained from the original distribution $\vec{P}$ with
\begin{eqnarray*}
\vec{P}^\prime=M_B\cdot\vec{P},\\
M_B=M_0^{-1}.
\end{eqnarray*}
For $n$-qubit simultaneous readout, $M_0$ is a $2^n\times{2^n}$ matrix. Each column of $M_0$ is a directly measured distribution vector along the corresponding basis. The experimental results of $M_0$ for the single-qubit and two-qubit measurements are:
\begin{eqnarray*}
M_{0,2\times{}2}=\left(\begin{array}{cc}
                   0.979 & 0.060 \\
                   0.021 & 0.940
                 \end{array}\right),\\
M_{0,4\times{}4}=\left(\begin{array}{cccc}
                   0.961 & 0.060 & 0.066 & 0.004 \\
                   0.024 & 0.925 & 0.002 & 0.065 \\
                   0.014 & 0.002 & 0.910 & 0.054 \\
                   0.001 & 0.013 & 0.023 & 0.877
                 \end{array}\right).\\
\end{eqnarray*}

\section{The Numerical Model of the System}
To simulate the quantum dynamics of the real device, we use the following Markovina master equation~\cite{RevModPhys.93.025005}
\begin{equation}
\dot{\rho}=-i[H(t),\rho]+\sum_q \gamma_{1q} \mathcal{L}[a_q]\rho+\sum_ q \gamma_{2q} \mathcal{L}[a^{\dagger}_q a_q]\rho
\label{master}
\end{equation}
that includes the environmental induced relaxation and dephasing effects. The Lindbladian $\mathcal{L}[a_q]$ represents the relaxation of the $q$-th qubit with rate $\gamma_{1q}=1/T_1q$, and the Lindbladian $\mathcal{L}[a_q^{\dagger}a_q]$ represents the pure dephasing of the $q$-th qubit with rate $\gamma_{2q}=1/T_{\phi{}q}$. Here $a_q$ is the annihilation operator of the $q$-th qubit, and the Lindbladian $\mathcal{L}[\mathcal{O}]\rho=\mathcal{O}\rho\mathcal{O}^\dagger-1/2\{\mathcal{O}^\dagger \mathcal{O},\rho\}$. The coherent part $H(t)$ reads:
\begin{equation}
\begin{aligned}
H(t)/\hhbar=& -\sum_q\frac{E_{C,q}}{2}a_q^\dagger{}a_q^\dagger{}a_qa_q+ \sum _{q>p}J_{qp} a_q^\dagger a_q a_p^\dagger a_p\\
& +\sum_q \left[ \theta_{2q-1}(t)(a_q^\dagger+a_q)/2+i \theta_{2q} (t)  (a_q-a_q^\dagger)/2\right ],
\end{aligned}
\end{equation}
where $E_{C,q}$ is the anharmonicity of the $q$-th qubit, and $J_{pq}$ is the dispersive coupling strength between qubits $p$ and $q$. The functions $\theta_{2q-1}(t)$ and $\theta_{2q}(t)$ are the control fields for the $q$-th qubit that induce $x$-axis and $y$-axis rotations, respectively.
%The relation between $\gamma_1$, $\gamma_2$ and $T_1$ (relaxation time), $T_2$ (coherence time) follows:
%\begin{equation}
%\begin{aligned}
%&\gamma_1=\frac{1}{T1},\
%&\gamma_2=\frac{1}{T2}-\frac{1}{2T1}.
%\end{aligned}
%\end{equation}

We use Eq.~(\ref{master}) to evaluate the influence of the relaxation and dephasing of the qubits on the learning performance, in which we assume all qubits have identical coherence times $T_1$ and $T_\phi$. Using the MNIST testing dataset, we examine the dependence of classification accuracies on the control pulse length, which are listed in Tables~\ref{T1} and \ref{T2}, respectively, under different $T_1$ and $T_\phi$. They collectively show that strong relaxation/dephasing tends to deteriorate the learning performance. Due to the voting method for assigning labels to input samples, an abrupt reduction to $50\%$ can be observed in the tables when the coherence times are too short, because the qubits quickly decohere to their equilibrium state no matter what data samples are fed in. To see the trends more clearly, we display in Fig.~3 of the main text the average confidence $1-\mathcal{L}$ evaluated on the same testing dataset, which are consistent with the tables listed here.

\begin{table}[]
    \caption{The classification accuracy under different pulse length $\tau$ and relaxation time $T_1$, while $T_\phi=\infty$. Here the pulse length includes all the $E+I=4$ sub-pulses in both the encoding and the inference blocks.}
    \label{T1}
    \begin{tabular}{|l|l|l|l|l|l|lll}
        \cline{1-6}
        \diagbox{$T_1$ ($\mu$s)}{length ($\mu$s)} & 0.08 & 0.16 &0.32  &0.64  & 1.28 &  &  &  \\ \cline{1-6}
        20                   & 0.981       & 0.983   &0.989  &0.987  &0.987  &  &  &  \\ \cline{1-6}
        10                   & 0.982       & 0.982   &0.988  &0.990  &0.987  &  &  &  \\ \cline{1-6}
        5                    & 0.981       & 0.983   &0.986  &0.986  &0.988  &  &  &  \\ \cline{1-6}
        2.5                  & 0.982      & 0.984    &0.988  &0.986  &0.981  &  &  &  \\ \cline{1-6}
        1.25                 & 0.982       & 0.986   &0.985  &0.984  &0.978  &  &  &  \\ \cline{1-6}
        0.625                & 0.982       & 0.985   &0.985  &0.986  &0.961  &  &  &  \\ \cline{1-6}
        0.3125               & 0.974       & 0.984   &0.981  &0.969  &0.500  &  &  &  \\ \cline{1-6}
        0.1563               & 0.973       & 0.977   &0.956  &0.500  &0.500  &  &  &  \\ \cline{1-6}
        0.0781               & 0.965       & 0.897   &0.500  &0.500  &0.500  &  &  &  \\ \cline{1-6}
        0.0391               & 0.500       & 0.500   &0.500  &0.500  &0.500  &  &  &  \\ \cline{1-6}
        0.0195               & 0.500       & 0.500   &0.500  &0.500  &0.500  &  &  &  \\ \cline{1-6}
    \end{tabular}
\end{table}

\begin{table}[]
    \caption{The classification accuracy under different pulse length $\tau$ and pure dephasing time $T_\phi$, while $T_1=\infty$. Here the pulse length includes all the $E+I=4$ sub-pulses in both the encoding and the inference blocks.}
    \label{T2}
    \begin{tabular}{|l|l|l|l|l|l|lll}
        \cline{1-6}
        \diagbox{$T_\phi$ ($\mu$s)}{length ($\mu$s)} & 0.08 & 0.16 &0.32  &0.64  & 1.28 &  &  &  \\ \cline{1-6}
        $\infty$                  & 0.982       & 0.983   &0.989  &0.989  &0.993  &  &  &  \\ \cline{1-6}
        20                   & 0.984       & 0.985   &0.990  &0.987  &0.994  &  &  &  \\ \cline{1-6}
        10                   & 0.983       & 0.985   &0.987  &0.987  &0.986  &  &  &  \\ \cline{1-6}
        5                    & 0.982       & 0.981   &0.987  &0.989  &0.978  &  &  &  \\ \cline{1-6}
        2.5                  & 0.982      & 0.983    &0.982  &0.984  &0.985  &  &  &  \\ \cline{1-6}
        1.25                 & 0.982       & 0.982   &0.984  &0.985  &0.978  &  &  &  \\ \cline{1-6}
        0.625                & 0.981       & 0.979   &0.982  &0.981  &0.982  &  &  &  \\ \cline{1-6}
        0.3125               & 0.981       & 0.978   &0.979  &0.978  &0.980  &  &  &  \\ \cline{1-6}
        0.1563               & 0.975       & 0.977   &0.980  &0.961  &0.955  &  &  &  \\ \cline{1-6}
        0.0781               & 0.971       & 0.961   &0.957  &0.945  &0.890  &  &  &  \\ \cline{1-6}
        0.0391               & 0.962       & 0.958   &0.940  &0.869  &0.724  &  &  &  \\ \cline{1-6}
        0.0195               & 0.952       & 0.934   &0.845  &0.695  &0.500  &  &  &  \\ \cline{1-6}
        0.0020               & 0.500       & 0.500   &0.500  &0.500  &0.500  &  &  &  \\ \cline{1-6}
    \end{tabular}
\end{table}

\section{{Linear Discriminant Analysis}}
The aim of linear discriminant analysis (LDA) is to project the high-dimensional data
vectors into two clusters of points distributed on an optimally
chosen line so that they can be better visualized.

Consider a dataset $\mathcal{D}=\{\vec{x}^{(k)},y^{(k)}\}$ with where $x^{(k)}\in\mathbb{R}^n$ and $y^{(k)}\in \{0,1\}$. Let $\mathcal{D}_0$ and $\mathcal{D}_1$ be the subsets associated with $y^{(k)}=0$ and $y^{(k)}=1$, respectively, the LDA analysis can be formulated as the seeking of a projector $\omega\in\mathbb{R}^n$ that maximizes
\begin{equation}
S(\omega)=\frac{\|\omega^\top (\mu_0-\mu_1)\|}{\omega^\top( \mathcal{E}_0 + \mathcal{E}_1)  \omega},
\end{equation}
where $\mu_c$ and $\mathcal{E}_c$ are the mean value and covariance matrix of $\mathcal{D}_c$, with $c=0,1$. The maximization of $S$ guarantees that the distance between the mean values of the projected points from the two clusters is the largest with the smallest variance in each cluster. It is easy to solve the optimal projector:
\begin{equation}
    \omega_{\rm opt}=(\mathcal{E}_0+\mathcal{E}_1)^{-1}(\mu_0-\mu_1), %\ \ \ \text{where}\ S=\mathcal{E}_0+\mathcal{E}_1.
\label{LDA}
\end{equation}
where the matrix inverse may be changed to pseudo-inverse when $\mathcal{E}_0+\mathcal{E}_1$ is singular.

The above LDA can be directly applied to the analysis of MNIST input data vector $\vec{x}$ and the converted encoding control variable $\vec{\theta}_\mathrm{En}$. As for the $D$-dimensiomal complex vectors $\ket{\psi(t_E)}$ and $\ket{\psi(t_{E+I})}$, we can transform them into $2D$-dimensional real vectors and apply the same LDA~\cite{Schlienz1995}. To facilitate the comparison of data points in different vector spaces, we further process the projected points after the LDA:
\begin{equation}
x^{(k)}=\frac{1}{||\omega_{\rm opt}^\top(\mu_0-\mu_1)||}\omega_{\rm opt}^\top \left(\vec{x}^{(k)}-\frac{\mu_0+\mu_1}{2}\right),
\end{equation}
which normalizes the distance between the centers of two data clusters and shifts the overall data center to the origin.

\section{{Algorithm for the classical optimizer}}
The algorithm for evaluating the loss function and the calculation of the gradient is shown in Algorithm~\ref{alg:gra_loss}. The Adam optimizer that is used in our experiment is shown in Algorithm~\ref{alg:adam}.

%\begin{algorithm}[H]
%\caption{Calculate the Gradient and the Loss Function of QNN}
%\label{alg:gra_loss}
%\begin{algorithmic}
%\State{$L_{out},g_W,g_{Infer}$ $\gets$ 0}
%\For{b=1:BatchSize}
%    \State{$L_0(b)$=Loss(1,b)}
%    \State{$g_0(:,b)$=(Loss(2:end,b)-$L_0$)/StepSize}
%    \State{$g_W=g_W+g_0$(IndexEncode)*xSet(:,b)$^T$/StepSize}
%    \State{$g_{Infer}=g_{Infer}+g_0$(IndexInfer)/StepSize}
%    \State{$L_{out}=L_{out}+L_0(b)$/StepSize}
%\EndFor
%\State \Return $L_{out},g_W,g_{Infer}$
%\end{algorithmic},
%\end{algorithm}

\begin{algorithm}[H]
\caption{Calculate the loss function and the gradients with respect to the model parameters}
\label{alg:gra_loss}
\begin{algorithmic}
		\State{\textbf{Input} $W,\vec{\theta}_{\text{In}},\text{Batch of training data}\{\vec{x}^{(k)},\vec{y}^{(k)}\}$}
		\State{$P_{\text{out}},\vec{g}_W,\vec{g}_{\text{In}}$ $\gets$ 0,  $\delta\in\mathbb{R}^{+}$}
		\For{$\ell=1:b$}
		 \State{$P_0=P(\vec{y}_\ell^{(k)}|\vec{x}_\ell^{(k)},W,\vec{\theta}_{\text{In}})$}
		\State{$\vec{\theta}_{\text{En}}=W*\vec{x}_\ell^{(k)}$}
		\For{$i=1:\text{Length}(\vec{\theta}_{\text{En}})$}
		 \State{$\vec{\theta}_1=[\vec{\theta}_{\text{En}}(1),\dots,\vec{\theta}_{\text{En}}(i)+\delta,\dots]$}
		 \State{$\vec{g}_1(i)=(P(\vec{y}_\ell^{(k)}|\vec{x}_\ell^{(k)},\vec{\theta}_1,\vec{\theta}_{\text{In}})-P_0)/\delta$}
		\EndFor
		\For{$i=1:\text{Length}(\vec{\theta}_{\text{In}})$}
		 \State{$\vec{\theta}_2=[\vec{\theta}_{\text{In}}(1),\dots,\vec{\theta}_{\text{In}}(i)+\delta,\dots]$}
		 \State{$\vec{g}_{2}(i)=(P(\vec{y}_\ell^{(k)}|\vec{x}_\ell^{(k)},\vec{W},\vec{\theta}_2)-P_0)/\delta$}
		\EndFor
		\State{$\vec{g}_W=\vec{g}_W-\vec{g}_1*\vec{x}_\ell^{(k)\top}/b$}
		\State{$\vec{g}_{\text{In}}=\vec{g}_{\text{In}}-\vec{g}_2/b$}
		\State{$P_{\text{out}}=P_{\text{out}}+P_0/b$}
		\EndFor
		\State \Return $P_{\text{out}},\vec{g}_W,\vec{g}_{\text{In}}$
	\end{algorithmic}
\end{algorithm}

%\begin{algorithm}[H]
%\caption{Adam optimizer}\label{alg:adam}
%\begin{algorithmic}
%\State{$lr=10^{-3}, \rho_1=0.9, \rho_2=0.999$}
%\If{$\text{iter}=1$}
%    \State{$s_1,s_2,r_1,r_2=0$}
%\EndIf
%
%\State{$s_1=\rho_1*s_1+(1-\rho_1)*g$}
%\State{$r_1=\rho_2*r_1+(1-\rho_2)*g^2$}
%\State{$s_{11}=s_1/(1-\rho_1^{\text{iter}})$}
%\State{$r_{11}=r_1/(1-\rho_2^{\text{iter}})$}
%\State{$p_{\text{iter}+1}=p_{\text{iter}}-lr*s_{11}/(\sqrt{r_{11}}+10^{-8})$}
%\State \Return $p_{\text{iter}+1},s_1,s_2,r_1,r_2$
%\end{algorithmic}
%\end{algorithm}

\begin{algorithm}[H]
\caption{Adam optimizer}\label{alg:adam}
\begin{algorithmic}
	\State{\textbf{Input}  $\text{Training dataset }\{\vec{x},\vec{y}\}$}
	\State{$\vec{s}_{w},\vec{r}_{w},\vec{s}_{\text{In}},\vec{r}_{\text{In}}\gets 0$}
    \State{$W,\vec{\theta}_{\text{In}}\gets W_0,\vec{\theta}_{\text{In},0}$}
    \State{$lr=10^{-3}, \beta_1=0.9, \beta_2=0.999,\epsilon=10^{-8}$}
	\For{$k=1:N$}
	\State{$\vec{g}_W, \vec{g}_{\text{In}}=\textbf{Algorithm1}(W,\vec{\theta}_{\text{In}},\{\vec{x}^{(k)},\vec{y}^{(k)}\}$)}
	\State{$\vec{s}_{w}=\beta_1*\vec{s}_{w}+(1-\beta_{1})*\vec{g}_W$}
	\State{$\vec{r}_{w}=\beta_{2}*\vec{r}_{w}+(1-\beta_{2})*\vec{g}_W.*\vec{g}_W$}
	\State{$\vec{s}_{1}=\vec{s}_{w}/(1-\beta_1^{k})$}
	\State{$\vec{r}_{1}=\vec{r}_{w}/(1-\beta_2^{k})$}
	\State{$W=W-lr*\vec{s}_{1}./(\sqrt{\vec{r}_{1}}+\epsilon)$}
	 \State{$\vec{s}_{\text{In}}=\beta_1*\vec{s}_{\text{In}}+(1-\beta_{1})*\vec{g}_{\text{In}}$}
	 \State{$\vec{r}_{\text{In}}=\beta_{2}*\vec{r}_{\text{In}}+(1-\beta_{2})*\vec{g}_{\text{In}}.*\vec{g}_{\text{In}}$}
	\State{$\vec{s}_{2}=\vec{s}_{\text{In}}/(1-\beta_1^{k})$}
	\State{$\vec{r}_{2}=\vec{r}_{\text{In}}/(1-\beta_2^{k})$}
	 \State{$\vec{\theta}_{\text{In}}=\vec{\theta}_{\text{In}}-lr*\vec{s}_{2}./(\sqrt{\vec{r}_{2}}+\epsilon)$}
	
	\EndFor
	\State \Return $W,\vec{\theta}_{\text{In}}$
\end{algorithmic}
\end{algorithm}

The gradient of the loss function $\mathcal{L}$ for parameter update in each iteration is obtained by averaging the gradients of the conditional probability $P$ over a batch of randomly selected input samples ($b=2$). This can reduce the fluctuation of $\mathcal{L}$ for faster convergence in the learning process. For the inference block, the gradient $\vec{g}_\text{In}$ of $\mathcal{L}$ with respect to each parameter in $\vec{\theta}_{\text{In}}$ is directly obtained by rerunning the experiment with a small change in $\vec{\theta}_{\text{In}}$ and calculating the difference of $P$. As for the encoding block, the gradient $\vec{g}_W$ with respect to the elements of the classical matrix $W$ is needed. To reduce the experimental cost, we can equivalently calculate $\vec{g}_W$ from the outer product between the measured gradient $\vec{g}_1$ with respect to the encoding controls and the input data vector.

Once the gradients with respect to the model parameters are obtained, we take Adaptive Moment Estimation (Adam)~\cite{Kingma2015} update algorithm to update the corresponding parameters. The Adam algorithm is popular for its efficiency and stability in stochastic optimization of learning problems. In the Adam algorithm, $lr,\beta_1,\beta_2,\epsilon$ refer to algorithm configuration parameters and are chosen as default empirical values. The intermediate parameters $\vec{s}_w,\vec{s}_{\text{In}},\vec{r}_w,\vec{r}_{\text{In}}$ will be passed to the next iteration so as to adaptively control the parameter updating rate.

%\bibliographystyle{Zou}
%\bibliography{citation}
%Control: production of eprint (-1) disabled
%